\documentclass[twocolumn]{biophys}
\usepackage{amssymb,amsfonts,amsthm,bm}
\usepackage[round,numbers,sort&compress]{natbib}
\usepackage[ansinew]{inputenc}
\usepackage{helvet,times}
\usepackage{bm,textcomp}
\usepackage{amsmath}
\usepackage{graphicx}

\jno{kxl014}
\gridframe{N}
\cropmark{N}
\doi{*mpz@nist.gov}
\volume{00}
\issue{0}
\Month{}
\Year{}
\usepackage{times}

\newcommand{\K}{K$^+$ }
\newcommand{\Cl}{Cl$^-$ }
\newcommand{\Na}{Na$^+$ }
\newcommand{\kt}{k_B T}

\newcommand{\avg}[1]{\langle #1 \rangle}

\newcommand{\pd}[2]{\frac{\partial #1}{\partial #2}}

\newcommand{\dFmax}{\Delta F^\star}

\begin{document}

\title{Optimal transport and colossal ionic mechano--conductance in graphene crown ethers}

\author
{Subin Sahu,$^{1,2}$ Justin Elenewski,$^{1,2}$ Christoph Rohmann,$^{1,2}$ Michael Zwolak$^{1\ast}$}

\address{$^{1}$Biophysics Group, Microsystems and Nanotechnology Division, Physical Measurement Laboratory, National Institute of Standards and Technology, Gaithersburg, MD 20899,$^{2}$Maryland Nanocenter, University of Maryland, College Park, MD 20742}

\begin{abstract}
{Biological ion channels balance electrostatic and dehydration effects to yield large ion selectivities alongside high transport rates.  These macromolecular systems are often interrogated through point mutations of their pore domain, limiting the scope of mechanistic studies. In contrast, we demonstrate that graphene crown ether pores afford a simple platform to directly investigate optimal ion transport conditions, i.e., maximum current densities and selectivity. Crown ethers are known for selective ion adsorption. When embedded in graphene, however, transport rates lie below the drift-diffusion limit. We show that small pore strains -- 1~\% -- give rise to a colossal -- 100~\% -- change in conductance.  This process is electromechanically tunable, with optimal transport in a primarily diffusive regime,  tending toward barrierless transport, as opposed to a knock--on mechanism. Measurements of mechanical current modulation will yield direct information on the local electrostatic conditions of the pore. These observations suggest a novel setup for nanofluidic devices while giving insight into the physical foundation of evolutionarily--optimized ion transport in biological pores.}
\end{abstract}

\maketitle
\markboth{Sahu et. al.}{Optimal transport and colossal ionic mechano--conductance}

\begin{figure*}
\includegraphics[width=\textwidth]{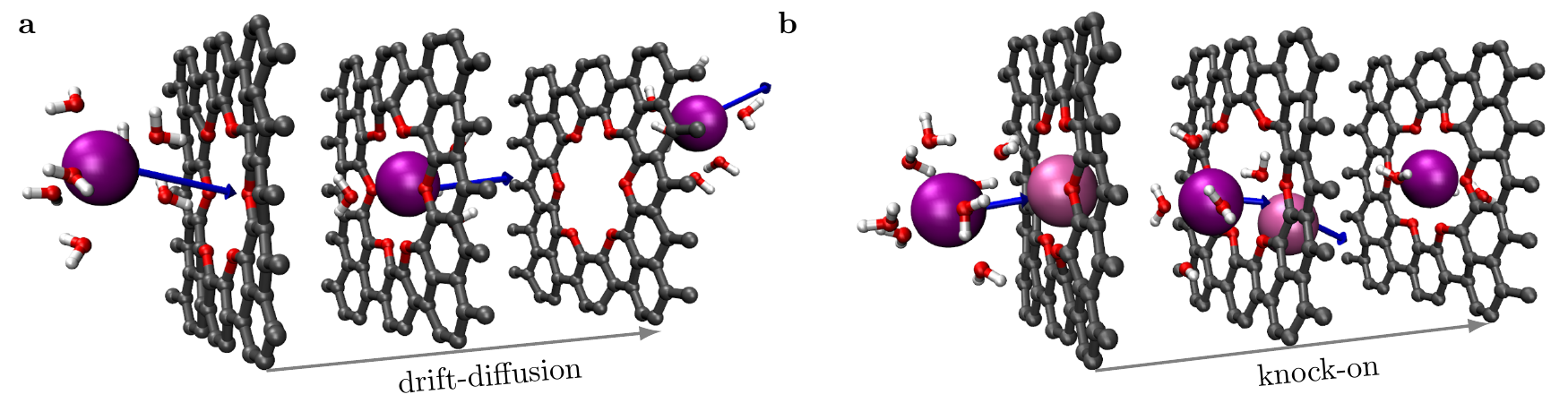}
\caption{{\bf Potential transport mechanisms in graphene crown ether pores.} {\bf a.} Using partial charges $q_\mathrm{O}=-0.24\,e$,  consistent with the electrostatic potential from DFT, the ion transport mechanism is drift--diffusion.  In this case, a \K (purple sphere) finds an empty pore and translocates through it; the pore then remains empty again for several nanoseconds (see the SM). {\bf b.} At larger partial charge ($q_\mathrm{O}=-0.54\,e$ or $q_\mathrm{O}=-1.0\,e$), the vicinity of the charge separation at the pore rim results in an energetic well deep enough to trap a \K (light purple). For a current to be present, an incoming \K (dark purple) knocks out and replaces the trapped K$^+$. This yields a two-step knock-on mechanism reminiscent of some biological potassium ion channels. However, for $q_\mathrm{O}=-0.54\,e$, the mechanism shifts towards a majority drift--diffusion process at moderate (1~\%) strain due to a shallowing of the free energy well. Oxygen, carbon, and hydrogen atoms are small red, grey, and white spheres, respectively.\label{Fig1}}
\end{figure*}

\section*{Introduction}

The search for universal transport mechanisms is an underlying theme in ion channel research~\cite{hille2001}. Nonetheless, even foundational questions -- such as the primary mechanism for selectivity of \K over Na$^+$ in the potassium ion channel family -- remain contentious: Is the smaller dehydration energy of the \K  ion responsible, or it is the ``snuggle--fit'' of \K  in the selectivity filter~\cite{nimigean2011origins, noskov2004control,Doyle98-1}? Solving these puzzles requires a detailed means to control and characterize ion transport.  Biological ion channels are notoriously difficult to manipulate, requiring mutagenesis and near--native conditions for patch--clamp studies, thus limiting experimental characterization and tunability.  These large macromolecules are also theoretically complex, requiring hefty resources to conduct statistically meaningful simulations. In contrast, biomimetic synthetic pores -- or, as addressed here, pores that retain core aspects of some biological channels -- may provide a simple platform to explore ion transport mechanisms under a broad range of conditions~\cite{kowalczyk2011biomimetic}.

In this work, we demonstrate that graphene crown ether pores~\cite{guo2014crown} have competing electrostatic and dehydration contributions to transport, reminiscent of biological ion channel mechanism and selectivity. The discovery of crown ethers a half--century ago triggered a cascade of research that evolved into the fields of host--guest and supramolecular chemistry~\cite{pedersen1967cyclic}. The isolation of graphene~\cite{novoselov2004electric} has likewise inspired diverse fields of study, including those that leverage atomically thin pores for biosensing and nanoscale separation~\cite{geim2010rise}. Graphene crown ether pores lie at the crossroads, affording a means to address transport and selectivity in a synthetic analog of biological ion channels. Using a graphene--embedded 18-crown-6 ether pore as a model system, we examine the mechanism for ion translocation under variable strain and applied voltage, see Fig.~\ref{Fig1}. Our results indicate that a few percent change in pore size can induce a several fold increase in ionic current, underscoring the sub--nanometer fine tuning between electrostatics and dehydration seen in biological contexts. We similarly find that small changes in pore radius and charge can optimize transport, and even alter the underlying mechanism of ion translocation. These results collectively suggest that graphene crown ether pores are an effective mechanistic probe for understanding transport in both biological and artificial ion channels alike.

\begin{figure*}[h]
\includegraphics[width=\textwidth]{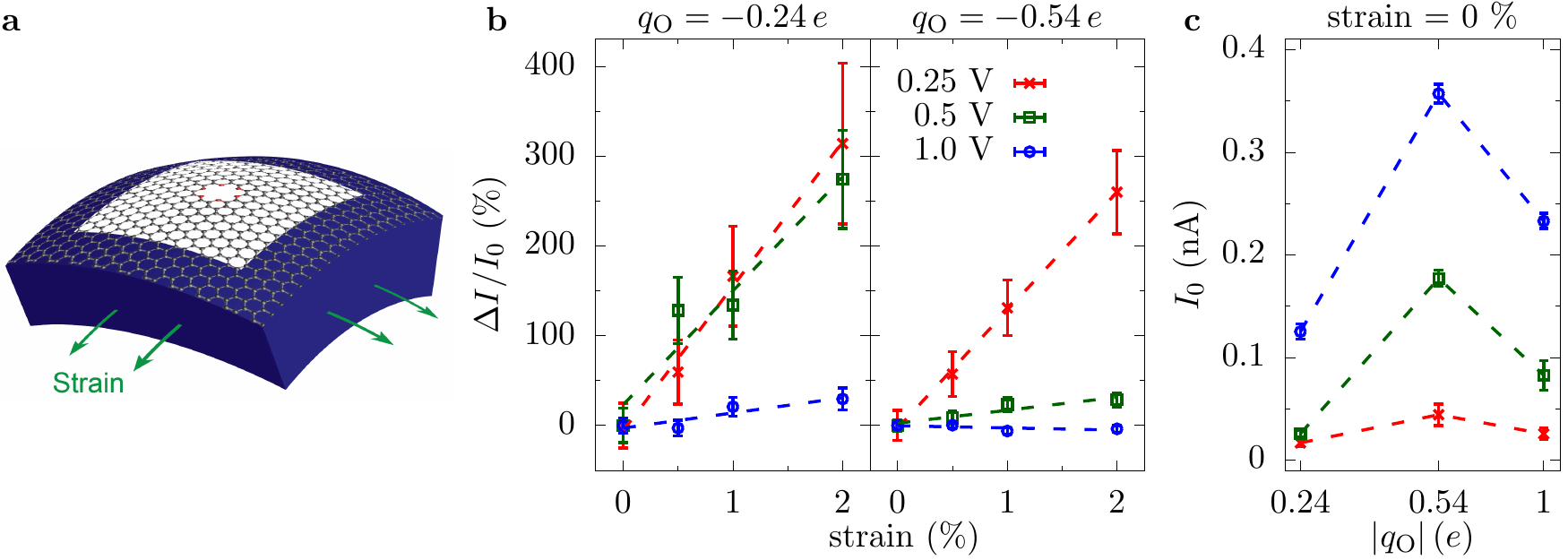}
\caption{{\bf Colossal ionic mechano--conductance.} {\bf a.} Schematic of graphene on (or embedded within) a polymeric matrix support (such as $\sim 50$ nm thick epoxy resin) with a window where the crown ether pore is located. The membrane can be strained by stretching or bending (e.g., via a piezoelectric actuator offset from the window). Alternative experimental setups are possible, such as metallic regions that bind the graphene and are used to apply strain. {\bf b.} Relative change in current versus strain in the crown ether pore (for $q_\mathrm{O}= -0.24\,e$ and $-0.54\,e$) at different voltages. At lower voltages, the current increases substantially with strain, as shown by the fitted dashed lines. {\bf c.} Current without strain ($I_0$) for different values of $q_\mathrm{O}$ and $V$. Going from $q_\mathrm{O}=-0.24\,e$ to $q_\mathrm{O}=-0.54\,e$, the current increases as highly charged oxygen atoms compensate the loss of waters from the hydration layers. However, going from $q_\mathrm{O}=-0.54\,e$ to $q_\mathrm{O}=-1.0\,e$, the current decreases because a further increase in the pore charge makes it harder for an ion to escape the potential well in the pore. The typical drift-diffusion limit -- set by access resistance to an uncharged pore with approximate radius 0.1 nm -- at 0.25 V is about 0.35 nA. Only the intermediate charge case -- the one most analogous to biological systems -- approaches this limit: It is about a factor of two lower at 2 \% strain. Further strain allows it to reach this limit. The smaller charge case can approach within a factor of 6. The range of accessible currents is due to the extensive mechanistic leeway permitted by the intermediate charge. The error bars are the standard error from five parallel runs.  \label{Fig2} }
\end{figure*}

\section*{Results}

\noindent{\bf Partial charge distribution:}
Unlike an isolated crown ether, an 18-crown-6 pore in graphene is believed to assume a planar conformation~\cite{guo2014crown}. However, while critical to molecular dynamics (MD) simulations, the charge distribution and response of this pore is unknown.  Prior simulations on this and related systems assume oxygen partial charges that span between $q_\mathrm{O} = -0.21\,e$ and $q_\mathrm{O} = -0.74\,e$~\cite{heath2018first, grootenhuis1989molecular, smolyanitsky2018aqueous, fang2019highly, wipff1982molecular, glendening1994ab}, over which transport properties will vary dramatically. To provide a more comprehensive approach, we reexamine this assignment using density functional theory (DFT), see the Methods and supplementary materials (SM) for details. Determinations from electrostatic potential fitting without an ion present -- consistent with additive force fields and the electrostatic environment predicted by DFT -- yield a uniform value of $q_\mathrm{O} = -0.24\,e$, decreasing to $-0.23\,e$ when a \K is present. Chemically distinct, but oxygen containing, fragments in OPLS and CHARMM force fields give substantially higher partial charges, $\ge -0.4\,e$, of which we take $q_\mathrm{O} = -0.54\,e$ as a representative example.  An upper bound of $q_\mathrm{O} = -1.0\,e$ is given by Bader analysis, which tends toward larger values of partial charge and represents the maximum expected from any electronic structure calculation. Physically reasonable values of $q_\mathrm{O}$ should span the low end of the range $-0.2\,e$ to $-0.6\,e$, as evidenced by electrostatic fits of $q_\mathrm{O} = -0.34\,e$ for the 18-crown-6 alkyl ether and $q_\mathrm{O} = -0.54\,e$ for a highly--dipolar glycine dipeptide backbone carbonyl. The graphene crown ether should be lower than both. We perform MD simulations using three distinct assignments $q_\mathrm{O}=(-0.24, -0.54, -1.0)\,e$ to capture the full range of potential behavior, showing both universal features of transport and aspects particular to different assignments. We test the effect of strain on ion transport using different applied biases and deformations, well within the elastic limit of graphene~\cite{pereira2009tight, lee2008measurement}.  \\

\noindent{\bf Ion transport mechanism:}
Since this pore is tiny and negatively charged, only \K contributes to the current in a KCl solution (set at a concentration of 1 mol/L in our case). The mechanism of transport depends on the ion's binding strength in the pore, largely regulated by $q_\mathrm{O}$ (as an indicator of the oxygen--carbon charge separation) and atomic structure. The mechanism is normal drift--diffusion for $q_\mathrm{O}=-0.24\,e$, shifting to a knock--on type for $-0.54\,e$ and $-1.0\,e$ (see the illustration in Fig.~\ref{Fig1}). For drift-diffusion, the residence time (on the order of 0.1~ns at 0.25 V) is much shorter than the delay to the next ion crossing event (on the order of 1~ns at 0.25 V). Conversely, for the two-step knock-on mechanism seen at higher charge, the residence time (on the order of 1 ns) is much larger than the delay time (less than 10 ps for most events; see the SM). However, even for $q_\mathrm{O}=-0.54\,e$,  drift--diffusion starts to dominate at high strain and/or voltage, both of which weaken ion binding to the pore.

\noindent{\bf Ionic mechano-conductance:}
Figure~\ref{Fig2}a depicts a potential experimental setup, in which a graphene sheet is selectively strained by deforming an underlying substrate.  Our simulations mimic these conditions by increasing the cross--sectional area of the membrane to induce a strain.  The relative change in current as a function of strain, for different values of $q_\mathrm{O}$ and applied voltages, is shown in Fig.~\ref{Fig2}b. At low voltages (0.25 V and 0.5 V), the ionic current increases rapidly with strain, orders of magnitude larger than expected from the increase in pore area alone.  Instead, this colossal increase is indicative of a change in the energetics and dynamics of ion transport. In the large voltage regime (1 V), the current remains constant with increasing strain when $q_\mathrm{O}= -0.54\,e$. This is not altogether surprising, as a large voltage drop across the membrane suppresses the influence of the free energy profile on translocating ions. The dependence of the current on $q_\mathrm{O}$ and $V$ is in Fig.~\ref{Fig2}c. In this case, the current increases with $q_\mathrm{O}$ between $q_\mathrm{O} = -0.24\,e$ to $q_\mathrm{O}= -0.54\,e$, while giving a turnover to decreasing current at higher charge.\\

\begin{figure}
\centering
\includegraphics{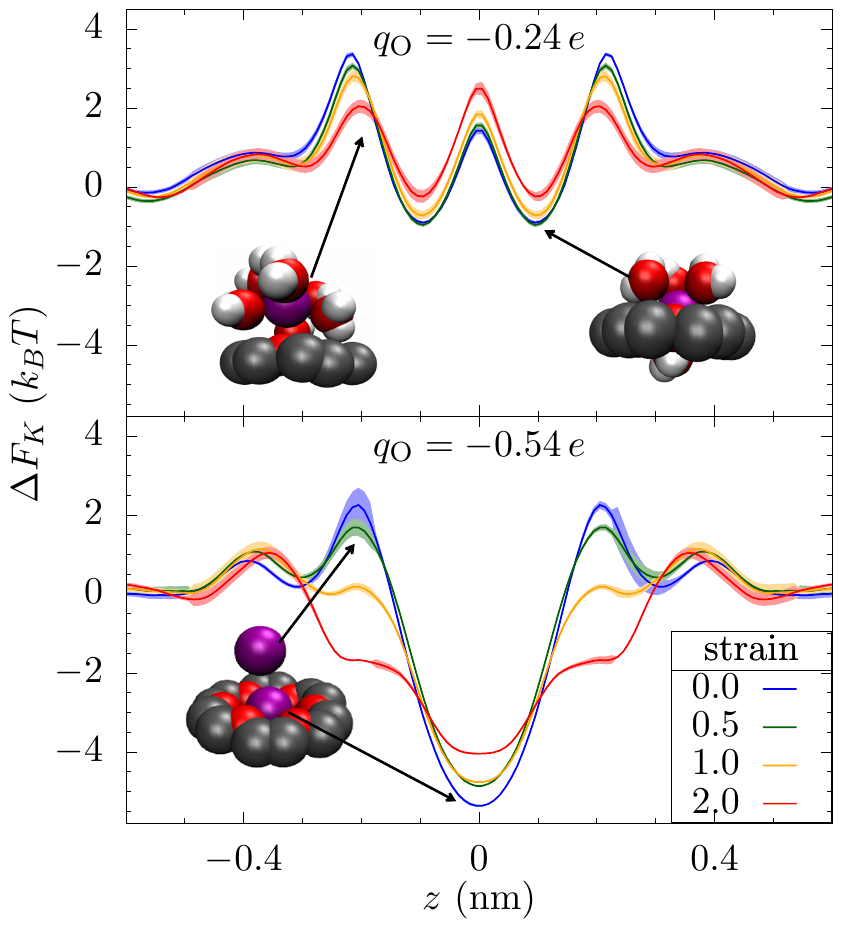}
\caption{ {\bf Free energy landscape.} Free energy profile, $\Delta F_\mathrm{K}$, for a \K translocating along the $z$-axis of an 18-crown-6 graphene pore at different strains. The charge on the oxygen atom, $q_\mathrm{O}$, is presented at the top of each plot. The peaks and valleys in $\Delta F_\mathrm{K}(z)$ are due to the balance between dehydration energy penalties and electrostatic interactions with the charged pore atoms. For $q_\mathrm{O}=-0.54$, there is an additional contribution from the \K  occupying the deep potential well, giving a barrier at $z\approx 0.2$ nm when the incoming ion attempts to go into the already occupied pore (this is confirmed by a free-energy profile with only 1 \K and 1 \Cl present; see the SM). The illustrations show the position of the \K  in the peaks and valleys. In general, the free energy profile flattens with increasing -- but still small --  strain, tending toward barrierless transport and making it easier for the ion to translocate. The error bands are the standard error from five parallel runs. \label{Fig3}}
\end{figure}

\noindent{\bf The energy landscape:}
Our results indicate that the energetics of pore permeation lie in a regime where small electromechanical variations modify current magnitudes and give rise to qualitative changes in behavior. The free energy of a potassium ion in the vicinity of the pore, $\Delta F_\mathrm{K}$, is approximately the sum of a dehydration energy barrier $\Delta E_\mathrm{deh}$ and three electrostatic terms ($E_\mathrm{KO}$, $E_\mathrm{KC}$, and $E_\mathrm{KK}$) that capture the interaction of \K with nearby charged atoms:
\begin{equation}\label{Eq:energy}
\Delta F_\mathrm{K} = \Delta E_\mathrm{deh}+E_\mathrm{KO}+E_\mathrm{KC}+E_\mathrm{KK},
\end{equation}
where $E_\mathrm{deh}=\eta \sum_i f_i E_i$ is the barrier due to dehydration, $f_i$ ($E_i$) is the fractional dehydration (energy) of the $i^\mathrm{th}$ hydration layer, and the factor $\eta\approx 1/2$ accounts for the nonlinear effects ~\cite{sahu2017dehydration}. Each electrostatic component is $E_{\mu\nu}=q_\mu n_\nu q_\nu / 4 \pi \epsilon_0 \epsilon(z_{\mu\nu}) r_{\mu\nu}$, where $\epsilon(z_{\mu\nu})$ is a position--dependent relative permittivity~\cite{gibby2018role}, $q_{\mu (\nu)}$ is the charge of ion species $\mu (\nu)$, $n_\nu$ is the number of proximate species $\nu$ (6 for oxygen, 12 for carbon, and 1 or 0 for a nearby K$^+$), $z_{\mu\nu}$ ($r_{\mu\nu}=\sqrt{z_{\mu\nu}^2+\rho_{\mu\nu}^2}$) is the axial (total) distance between species $\mu$ and $\nu$, and $\rho_{\mu\nu}$ is the respective radial distance. $E_\mathrm{KK}$ is the energetic coupling of an incoming \K to a trapped K$^+$. The van der Waals interactions, the coupling to other solvated ions, and the entropy change should have only marginal contributions when $\Delta F_\mathrm{K}$ varies with strain (e.g., the entropy change into the pore is important, but this varies little with moderate strain). In order to quantify the variable electrostatic environment reflected in $\epsilon(z)$, as well as nonlinear effects in $\Delta E_\mathrm{deh}$ (stronger polarization of bound water molecules as an ion becomes dehydrated~\cite{sahu2017dehydration}), we employ all--atom MD simulations to find the free energy profile of \K around the pore (Fig.~\ref{Fig3} and the Methods). Equation~\ref{Eq:energy} explains the key features in the free energy profiles for the range of $q_\mathrm{O}$ under consideration. Together with the MD data, this expression dissects the contributions to the free energy. We will use this equation below to characterize the response of ion transport to strain.

Figure~\ref{Fig3} shows that when $q_\mathrm{O}=-0.24\,e$,  a free energy barrier appears outside the pore due to the progressive dehydration of K$^+$. The barrier decreases with increasing proximity to the pore opening as the flanking oxygen atoms begin to compensate for water molecules displaced from the ion's first hydration layer. This results in a potential well on both sides of the membrane. Further dehydration occurs as the ion enters the pore, giving rise to an overall barrier at $z=0$ since the electrostatic attraction between the pore oxygen atoms and the \K  ($E_\mathrm{KO}$)  is insufficient to compensate for the dehydration energy penalty. This central barrier -- and only shallow adjacent wells -- ensure that an approaching \K will almost always find an empty pore, and hence the $E_\mathrm{KK}$ term will not contribute in this case. Equation~\ref{Eq:energy} confirms this assignment of the peaks and wells.

This situation is inverted for $q_\mathrm{O}=-0.54\,e$, see Fig.~\ref{Fig3}, where a deep potential well forms at the center of the pore due to strong electrostatic attraction with the K$^+$. The central pore acts as a binding site in this case, ensuring that it is almost always occupied. When \K approaches the pore, the presence of an interstitial \K  repels it, yielding a potential barrier on either side of the pore. A free energy calculation with single \K and single \Cl confirms this (see the SM). We can estimate $\epsilon_r(z)$ using the free energy profile from Fig.~\ref{Fig3} and the fractional dehydration from MD. For the physical range of pore charge, $\epsilon_r(z)$ is around 4 to 6 within 0.2~nm of the pore, which is consistent with experimental values for the dielectric constant within 1 nm of an interface~\cite{fumagalli2018anomalously}. A similar value for dielectric constant is given in another recent computational study~\cite{gibby2018role}.

\noindent{\bf Barrierless transport:}
While the transport mechanism is different for unstrained pores, the colossal mechano--conductance change is present at low voltage regardless of the partial charge. In both partial charge scenarios, the free energy profile becomes smoother with increasing strain, tending toward ``barrierless transport''~\cite{fedorenko2018quantized, gibby2018role}: Pulling charged oxygens away from the pore center reduces their coupling to translocating ions. The colossal change in conductance indicates that there are optimal structural positions for the oxygens: Picometer changes in their position do not change the dehydration contribution to the central barrier (see the SM) but it vastly changes the transport rate, \textit{maintaining exclusion of other ions but enhancing transport rates by several fold}. Biological ion channels can further make use of the partial charge, engineering not only structural characteristics, but also the electronic environment. As we discussed earlier, the larger partial charges we consider -- in line with biological channels~\cite{Bucher2009} --  show a shift from knock--on to drift--diffusion mechanisms as transport starts to become barrierless. In this manner, the electromechanical environment may exert a large effect on the current.\\

\noindent{\bf Mechanical susceptibility:}
To understand the mechanics underlying this process, we define the susceptibility, $\chi$, of the free energy to a small change in pore size,
\begin{equation} \label{eq:chi}
\chi = \frac{d \Delta F}{ d a} = \sum_{\avg{\mu \nu}}  \pd{\Delta F}{\rho_{\mu \nu}} \frac{d\rho_{\mu \nu}}{da} \approx  q_\mathrm{K} E_\rho ,
\end{equation}
where $a$ is the effective pore radius, $\avg{\mu \nu}$ indicates the pairs KO and KC (we ignore the role of KK interactions, although this is important for knock--on aspects of transport), and $d\rho_{\mu \nu}/da \approx 1$.  The latter approximation is reasonable since, as the pore size increases, the oxygen and carbons move outward by the same amount. For simplicity, we took this for an ion at the origin ($z=0,\rho=0$) where the susceptibility is just the \textit{local radial electric field}, $E_\rho$, multiplied by the charge of the translocating ion. Due to the lack of electrostatic screening in the pore (i.e., $\epsilon(z=0)$ is small) and the proximity of the oxygen atoms, this field is very large, resulting in a large $\chi$: Equation~\eqref{Eq:energy} gives $\chi\approx 100\, \kt$/nm  and $\chi\approx 140\, \kt$/nm for the $q_\mathrm{O}=-0.24\,e$ and $q_\mathrm{O}=-0.54\,e$, respectively (using $\epsilon=3.9$ and $6.4$, which we extract via Eq.~\eqref{Eq:energy} and the all-atom MD data). These two susceptibilities are closer than their difference in charge would indicate due to the smaller permittivity for $q_\mathrm{O}=-0.24\,e$ (the charge $q_\mathrm{O}=-0.54\,e$ more strongly attracts counter charges, specifically \K and hydrogens on water molecules, that help screen the interaction.)

\textit{The large susceptibility means that even a one picometer change in the pore size gives a 0.1~$\kt$ to 0.14 $\kt$ change in the free energy}. A 1~\% strain gives about 7 pm change in radius. Using this in Eq.~\eqref{Eq:energy}, free energy changes by about 0.7~$\kt$ for $q_\mathrm{O}=-0.24\,e$ and about 1.0~$\kt$ for  $q_\mathrm{O}=-0.54\,e$, in line with the all-atom results in Fig.~\ref{Fig3}. We note that there are clearly other effects occurring: As we will discuss below, the $q_\mathrm{O}=-0.24\,e$ case is already in a limit where dehydration and electrostatic effects are comparable, and thus the change in the free energy profile versus $z$ reflects both (the barrier reduction at $z\approx 2$~nm is due to a changing dehydration contribution, whereas in between those two outer barriers, electrostatic effects dominate and $\Delta F$ increases overall). For  $q_\mathrm{O}=-0.54\,e$, the potential well shallows, causing a transition from a knock--on to a diffusive mechanism and removing the ion--ion repulsion that causes the external barrier. Nevertheless, in both cases, the electrostatic contribution to the susceptibility captures the increased free energy for an ion in the pore. While a general susceptibility will need to include the totality of effects, electrostatics and dehydration will typically dominate and are key to understanding amplification. 

The free energy profile and its variation (according to $\chi$) determines the current. In particular, the largest free energy hurdle, $\dFmax$, in the profiles of Fig.~\ref{Fig3} are the main factors influencing the conductance. For $q_\mathrm{O}=-0.24\,e$, this will be the jump from about $z=$ 0.1~nm to 0.2~nm and, for $q_\mathrm{O}=-0.54\,e$, this will be the jump out of the well ($z=0$) to the maxima (at about $z=0.2$~nm to $z=0.4$~nm depending on strain). The current will take the form (see Methods) 
\begin{equation}\label{Eq:J}
I \approx A e^{-\dFmax/\kt},
\end{equation}
where $A$ is a constant that only weakly depends on pore parameters (such as size), but we will take it as fixed. The total amplification of the current then becomes 
\begin{equation}\label{eq:amp}
\frac{\Delta I}{I} \approx e^{(-\dFmax(s) + \dFmax(s=0))/\kt}-1 \approx  e^{\alpha\,a\,s\,\chi/\kt} - 1,
\end{equation}
where $-\dFmax(s) + \dFmax(s=0) \approx  d \Delta F /da \cdot \delta a \approx \alpha\,a\,\,s\,\chi$. The parameter $\alpha = 1/a \cdot da/ds$ quantifies the relative response of the pore radius to the application of a strain. Both MD and DFT predict that $\alpha \approx 2$, meaning that if the graphene has 1~\% strain, the atoms at the pore rim move by about 2~\%. This is due to a distortion of the rim structure since the pore is an effective impurity and can relieve strain by relaxing angular coordinates. The straightforward relation in Eq.~\ref{eq:amp} indicates that the gain in current is the exponential of the work performed by the applied strain on a \K within the pore. For the values in our case, Eq.~\eqref{eq:amp} gives an amplification of  100~\% to 200~\% for a 1~\% strain, in line with the MD results in Fig.~\ref{Fig2}b. \\

\begin{figure}
\centering
\includegraphics{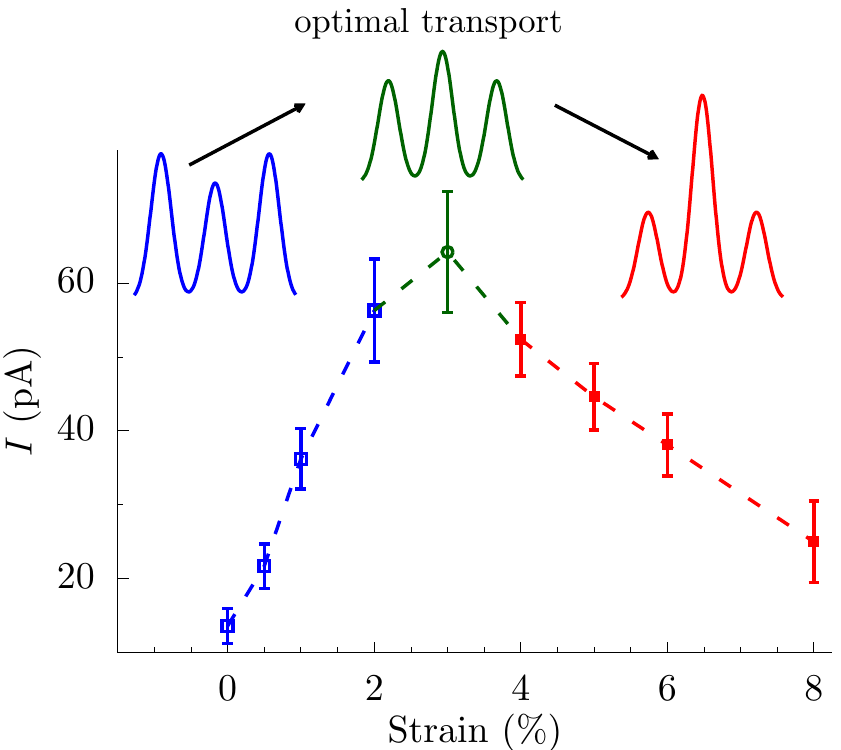}
\caption{{\bf Optimum ion transport and selectivity}. Ionic current versus strain in a pore with $q_\mathrm{O}=-0.24\,e$ and an applied bias of 0.25~V. For small strain (blue line), the current increases rapidly, commensurate with a decrease in the outer barriers with increasing strain. This gives an overall flattening of the free-energy profile.  Transport (and selectivity, see below) is optimal near 3~\% strain (green line) when the outer barrier is sufficiently diminished but the central barrier is not too high.  Further increases in strain continue to increase the central barrier (dehydration remains unchanged in this strain regime), decreasing the current (red line). Schematics of the free-energy profiles are above each regime.  For every value of strain and voltage, the selectivity of \K over \Cl is perfect for all practical purposes, as the negative partial charge of the oxygens on the pore rim create a strong barrier for \Cl transport. Separate simulations (each of duration  1 $\mu$s with an equal mixture of NaCl and KCl at 1 mol/L Cl$^-$) reveal that  \Na does not cross the pore (except for 0~\% strain where a single Na$^+$ crossing event occurs during the 1~$\mu$s time frame). Due to the near complete exclusion of \Na in the time frame of the simulation, we can determine a high-confidence lower bound on the selectivity, which goes from 15 at 0~\% strain to 52 at 3~\% strain, see the text for details. In other words, the \Na and \Cl exclusion is maintained while \K current changes with strain, and the lower bound on selectivity exactly traces the \K current in the plot. 
\label{Fig4} }
\end{figure}

\noindent{\bf Optimum transport:}
In all parameter regimes, the amplified current is due to an overall flattening of the free energy profile. Ion transport thus tends toward barrierless transport for small values of strain and displays a colossal increase in conductance along the way. However, for $q_\mathrm{O}=-0.24\,e$, the dominant contribution to $\dFmax$ shifts from a pathway connecting a minimum (at $z=\pm\,0.1$~nm) and the external solution to a pathway that hops over the central barrier (the well minima already lie at about $\Delta F = 0$). After this point, $\dFmax$ will continually increase due to decreasing compensation by the oxygen partial charge. There will be a corresponding decrease in current, see Fig.~\ref{Fig4}, thus indicating an optimum near this value of strain -- a regime that may be experimentally accessible.

In separate calculations with an equal mixture of NaCl and KCl (with \Cl at one molar), we find that only a single Na$^+$ traversed the pore (compare to  47 K$^+$) during 1 $\mu$s of simulation for the unstrained pore at a 0.25 V bias. At higher strains, only K$^+$ translocate through the pore within the 1 $\mu$s simulation time. The lack of \Na transport is due to the larger hydration energy of \Na compared to K$^+$, giving it a higher penalty to go through the pore. The presence of an overall barrier should result in \Na crossings following Poisson statistics. Due to rarity of these crossings, we consider two values for the \Na current, 0.5 pA and 0.16 pA . The former would give a 5~\% probability to detect no \Na crossings during 1 $\mu$s and thus is a high confidence \textit{upper bound} to the \Na current. The latter gives equal probabilities for 0 and 1 \Na crossing and thus represents an estimate of the \Na current assuming we are near the threshold to start seeing \Na crossings, for which the single crossing at 0~\% strain suggests is the case. Using the \textit{upper bound} on the \Na current, the corresponding high-confidence \textit{lower bound} on the selectivity ranges from about 15 at 0~\% strain to 52 at 3~\% strain (note that the \K currents in the equal mixtures are 7.5 pA at 0~\% strain and 26.2 pA at 3~\% strain, which are approximately half the value in the 1 mol/L KCl case shown in Fig.~\ref{Fig4} due to the mixture having half the \K concentration). The estimate of selectivity, though, ranges from 47 at 0~\% strain to 164 at 3~\% strain. Very long simulations (much greater than 10 $\mu$s) will be necessary to tightly bound the selectivity, but it will be large nonetheless.

For $q_\mathrm{O}=-0.54\,e$, the free energy at $z=0$ will not turn from a well to barrier until about $s=6$~\% (i.e., about $\delta a = 6\,\kt/\chi$), with the optimum appearing at potentially larger strains and outside of measurable regimes. Up to this point, the current will increase with strain as the free energy profile becomes truly barrierless. In either case, the dehydration contribution itself (at $z=0$) will not substantially change until the pore becomes sizable, as indicated by the continuing decrease of the current in Fig.~\ref{Fig4}. The selectivity filters of biological channels have partial charges near $q_\mathrm{O}=-0.54\,e$  by virtue of the dipolar peptide backbone~\cite{Doyle98-1,Roux2005, Kuang2015, Bucher2009}, lying within this more strongly coupled regime and affording them extensive mechanistic leeway.  These systems generically exploit the spatial distribution of charges or functional groups via the protein structure to achieve optimality. Strain in the graphene crown ether pore modulates a single collective variable, acting as a model for this optimization process.

\section*{Discussion}
The investigation of 18-crown-6 and other crown ether pores, as well as pores generally, with respect to strain (the experimental extraction of $\chi$ and determination of the local interaction environment), voltage, and ion concentration will shed further light into the behavior of selectivity and optimal transport. These characteristics underlie the mechanism of physiological ion channels and are of substantial interest in biology and medicine~\cite{hille2001,kandel2000}. This is paralleled by broad applicability to technologies for the sensing and sequencing of biomolecules~\cite{deamer2016three, sahu2019colloquium, prozorovska2018state}; 
 molecular sieving and desalination~\cite{Joshi2014, park2017maximizing}; 
 as well as batteries, fuel cells, and other energy harvesting devices~\cite{wang2011graphene,siria2013giant}. In all of these applications, the atomic thickness and physical strength of graphene provide a unique advantage~\cite{prozorovska2018state, sahu2019colloquium} facilitating the detailed examination of molecular processes such ion dehydration~\cite{sahu2017dehydration, sahu2017ionic}. Mechanical modulation of ion transport will thus lend itself to filtration and other technologies. 

Our results indicate the ideal pore characteristics for optimal transport: Minute changes in pore size or charge give rise to a dramatic change in ionic conductance as transport becomes barrierless, but further change will decrease the conductance. These systems also comprise a simple and transparent platform to understand the evolution of biological ion channels.  Working by analogy, it is possible to identify structural features and charge distributions that maximize selectivity while maintaining high permeation rates.  These considerations, in turn, constrain which conformational transitions may be leveraged to regulate transport~\cite{Baker2007}.  Layered crown ether graphene pore heterostructures -- mimicking aspects of the selectivity filter in the potassium channel from \textit{Streptomyces lividans} (KcsA) -- are likewise promising for workable models of biomolecular channel dynamics.  These systems should be quite sensitive to ionic conditions, potentially gating or doping graphene to access broad regimes of behavior. The susceptibility to transverse strain is also generally interesting for artificial porous membranes as it gives otherwise inaccessible information regarding the local electrostatic and hydration conditions. This platform will thus open new and general opportunities for the study, quantification, and tuning of ion transport at the nanoscale.

\section*{Methods}

We determine point charges using both localized--basis and extended plane--wave DFT schemes.  The localized basis calculations (B3LYP/6-31G*) employ a truncated model of the graphene pore alongside the Gaussian09 code~\cite{g09}, affording point charges through CHELPG electrostatic potential fitting \cite{Breneman1990}.  In contrast, for plane--wave DFT, we employ VASP\cite{Kresse1996b}  and a dispersion--corrected scheme (PBE-D2/PAW) for a 1.71 nm $\times$ 1.73 nm graphene supercell containing an 18-crown-6 pore, with periodic images  separated by a 2.0 nm vacuum layer ($5 \times 5 \times 1$ Monkhorst--Pack mesh, 500 eV cutoff) \cite{monkhorst_special_1976}. In this case, point charges were determined using Bader analysis \cite{tang_grid-based_2009}. A complementary set of plane wave calculations employs a similar pore within a 2.96 nm $\times$ 2.99 nm cell, described using CP2K \cite{Hutter2014} and a mixed Gaussian plane--wave basis scheme (PBE-D3/GPW/DZVP/GTH pseudopotentials, 5442 eV real--space grid cutoff) 
with $\Gamma$-point sampling and RESP charge fitting \cite{Bayly1993}.  
  Point charges are derived from optimized geometries, defined where SCF energy convergence is below $1.0 \times 10^{-4}$ eV and forces below 0.1 eV nm$^{-1}$. CHELPG fitting yields $q_\mathrm{O}=-0.24\,e$ without a \K present in the pore and $-0.23\,e$ when a \K is present. The complementary plane-wave calculations confirm these assignments. The Bader analysis yields  $q_\mathrm{O}=-1\,e$, an upper bound for local partial charges due to its strong bias toward chemically intuitive, but not electrostatically representative, distributions. Both MD (below) and DFT give an $\alpha$ of about 2 (DFT yields a slightly higher value), indicating that the pore rim has an enhanced response to the strain. Reference fragments of $q_\mathrm{O} = -0.54\,e$ include calculated CHELPG carbonyl charges for the Ac-Gly-Gly-CH$_3$ peptide (B3LYP/6-31G*), as well as standard OPLS ($q_\mathrm{O} = -0.50\,e$) and CHARMM ($q_\mathrm{O} = -0.51\,e$) backbone carbonyl parameters. Graphene should have oxygen partial charges lower than any of these fragments' values -- as well as the isolated crown ether's oxygen partial charges -- due to the delocalized nature of the electrons in the carbon sheet (see the SM for details and extended methods).

We simulate ion transport through the pore using all--atom molecular dynamics via the NAMD2 code~\cite{phillips2005} with an integration time step of 1 fs. We take a simulation cell aspect ratio of 1.2 -- the \textit{golden aspect ratio} -- where the simulation captures both access and pore resistances~\cite{sahu2018maxwell, sahu2018golden}. While access resistance should be less than a 10~\% correction for some parameter regimes, it must be taken into account when understanding how the drift--diffusion limit is approached. Its exclusion by choosing inappropriate (either too small or a poor aspect ratio) simulation cells hinders the comparison with the drift--diffusion limit (also see the SM). 

We employ the adaptive biasing force (ABF) method~\cite{Henin2004} to calculate the equilibrium free energy $\Delta F$ profile for a \K going through the pore. We compute $\Delta F$ versus $z$ in a cylindrical region of radius 0.1 nm and length 3 nm centered around the pore. As a \K approaches the pore, it experiences either a free energy barrier or a potential well. Assuming a small applied voltage $V$, the current over a barrier $\Delta F > 0$ is~\cite{nelson2002permeation}
\begin{align}\label{I_Barrier}
I_\mathrm{K}&=2\,q_\mathrm{K}\, k_\text{in}\, e^{-\Delta F/ \kt} \sinh(q_\mathrm{K}V/2\kt), 
\end{align}
and the current across a potential well  $\Delta F < 0$ is 
\begin{align}\label{I_Well}
I_\mathrm{K}&= \frac{2\,q_\mathrm{K}\, k_\text{in}\,k_\text{out}\,c\, e^{-\Delta F/ \kt} \sinh(q_\mathrm{K} V/2\kt)}{k_\text{in}\, c+ k_\text{out}\, e^{-\Delta F/\kt}\, \cosh(q_\mathrm{K}V/2\kt)}, \nonumber\\
\end{align}
where $k_\text{in}$ and $k_\text{out}$ are the rate constant for incoming and outgoing ions and $c$ is the ion concentration in bulk solvent. Both of these expressions give Eq.~\ref{Eq:J} for large $\Delta F$. (See the SM for a additional discussion and results regarding the influence of the free energy barrier, including a multiple barrier rate model).

\section*{Acknowledgments}
S. Sahu, J. Elenewski, and C. Rohmann acknowledge support under the Cooperative Research Agreement between the University of Maryland and the National Institute of Standards and Technology Center for Nanoscale Science and Technology, Award  70NANB14H209, through the University of Maryland. 

\noindent{\bf Author contributions:} S.S. analyzed the ion transport behavior and performed MD simulations. J.E. and C.R. performed DFT calculations. S.S., J.E., and M.Z. formulated and derived mathematical expressions. All authors developed the ideas and wrote the manuscript. 

\noindent{\bf Competing interests:} The authors declare that they have no competing interests.

\noindent{\bf Data and materials availability:} All data necessary to evaluate the conclusions in the paper are present in the paper and/or the Supplementary Material. Additional data related to this paper may be requested from the authors.

%\onecolumn
%\bibliography{ref}

\end{document}

\title{Optimal transport and colossal ionic mechano--conductance in graphene crown ethers --- Supplementary Materials}

\author
{Subin Sahu,$^{1,2}$ Justin Elenewski,$^{1,2}$ Christoph Rohmann,$^{1,2}$ Michael Zwolak$^{1\ast}$}

\address{$^{1}$Biophysics Group, Microsystems and Nanotechnology Division, Physical Measurement Laboratory, National Institute of Standards and Technology, Gaithersburg, MD 20899,$^{2}$Maryland Nanocenter, University of Maryland, College Park, MD 20742}

\maketitle 

\markboth{Sahu et. al.}{Optimal transport and colossal ionic mechano--conductance}

\tableofcontents
%\clearpage

\vspace{2\baselineskip}
\section{Electronic structure calculations}

We employ electronic structure calculations, including both localized--basis and plane--wave density functional theory (DFT), to determine the lattice parameters of graphene--embedded 18-crown-6 pores, alongside point charge distributions for molecular dynamics simulations.

\subsection{Mechanical response and Bader charge determination}

We first examine the mechanical and electronic properties of 18-crown-6 pores using periodic DFT calculations using a projector augmented wave (PAW) basis and the Perdew--Burke--Ernzerhof (PBE) density functional, alongside Grimme's D2 dispersion correction~\cite{perdew_generalized_1996,grimme_stefan_semiempirical_2006,kresse_ultrasoft_1999}, all within the Vienna \textit{Ab initio} Simulation Package (VASP)~\cite{Kresse1996b}.  We allow atoms in the pore to relax until the total energy converges to below $10^{-4}$ eV between successive SCF steps and until the forces on each atom are less than 0.1 eV/nm. A 2 nm vacuum layer above and below the pore prevents interaction between periodic slab images. We sample the Brillouin zone with a $5\times 5\times 1$ Monkhorst--Pack mesh \cite{monkhorst_special_1976} and use an energy cut-off of 500 eV for these calculations.

Lattice parameters ($a=b=0.2468$ nm) are derived a fully--relaxed $7\times7$ graphene sheet,  in excellent agreement with values ($a=b=0.246$ nm) commonly reported in the literature~\cite{meunier2016physical}. Using these parameters, we construct an 18-crown-6 pore in a rectangular graphene supercell (1.7099 nm $\times$ 1.7276 nm) and optimize it for each value of strain. A strain of 0.5~\%, 1.0 \%, and 2.0 \% will result in the pore opening by 1.1 \%, 2.2 \% and 4.4 \%, respectively. This factor of approximately two in response occurs due to an increase in the C--O--C angle from about 120.8$^\circ$ to 123.7$^\circ$ (for the case of 2.0 \% strain).  Our data shows that the changes in the local C--O and local C--C bond lengths are commensurate with the magnitude of the strain, see Table~\ref{tab:VASP}. Nonetheless, our results show that the opening of the pore is accompanied by a slight decrease in the charge on the O and C atoms, becoming less negatively and positively charged, respectively. This will not change the colossal mechano-conductance at small strain regardless of the pore charge. It will, though, reinforce the turnover behavior at large strain, as the electrostatic compensation for dehydration will lessen as the pore opens further.

We also calculate point charges in this system via Bader analysis~\cite{tang_grid-based_2009}. While not consistent with electrostatic potential distributions, the Bader charges represent a chemically intuitive distribution of localized (``classical'') charge, see Table~\ref{tab:VASP} and~\ref{tab:charges}.

\begin{table} [h]
\centering
\def\arraystretch{1.2}
\setlength\tabcolsep{8 pt}
\begin{tabular}{|l|c|c|c|c|}
\hline 
Strain & 0\% & 0.5\% & 1\% & 2\%\tabularnewline
\hline 
\hline 
{$d$(C--C)$^{*}$ (nm)} & {0.141} & {0.142} & {0.143} & {0.145}\tabularnewline
\hline 
{$d$(C--O)$^{**}$ (nm)} & {0.140} & {0.140} & {0.141} & {0.142}\tabularnewline
\hline 
{$\theta$(C--O--C) (deg.)} & {120.8} & {121.6} & {122.3} & {123.7}\tabularnewline
\hline 
{$d$(C, pore centroid) (nm)} & {0.372} & {0.375} & {0.378} & {0.384}\tabularnewline
\hline 
{$d$(O, pore centroid) (nm)} & {0.283} & {0.286} & {0.289} & {0.295}\tabularnewline
\hline 
{$q_\text{O}$} & {-1.05\,$e$} & {-1.04\,$e$} & {-0.96\,$e$} & {-0.93\,$e$}\tabularnewline
\hline 
{$q_\text{C}$} & {0.48\,$e$} & {0.47\,$e$} & {0.45\,$e$} & {0.41\,$e$}\tabularnewline
\hline 
\end{tabular}
\caption{Geometric parameters and Bader charges from plane--wave DFT. These parameters include the bond lengths and distances $d$, angles $\theta$, and Bader charge assignments $q_\mu$ ($\mu=\mathrm{O}$ or C) for an 18-crown-6 pore in graphene. The C--C bond lengths (*) are an average over only those carbon atoms directly flanking the pore and the C--O bond lengths (**) are an average over all such pairs in the pore.\label{tab:VASP}}
\end{table}

\subsection{Electrostatic Potential Fitting: Localized Basis}

To maintain consistency with the additive CHARMM force field~\cite{Feller2000}, we determine oxygen point charges for MD using DFT in a localized Gaussian basis (B3LYP/6-31G{*}) via the Gaussian09 code~\cite{Hariharan1973, Lee1988,Becke1993,Stephens1994,Hariharan1973,g09}. Calculations employ a truncated model of the graphene pore, Fig.~\ref{fig:DFTpointcharges}a. Electrostatic potential fitting uses the CHELPG scheme~\cite{Breneman1990}, see Table~\ref{tab:charges}. Similar results are given by CHELPG/RI-MP2/aug-cc-pVDZ charges on top of an RI-MP2/cc-pVDZ geometry from a simpler model \cite{Dunning1989,Kendall1992}, Fig.~\ref{fig:DFTpointcharges}b. We also examine a highly polar oxygen group -- specifically the backbone carbonyl of a glycine--glycine dipeptide -- common in biological channels via CHELPG/6-31G$^*$/B3LYP.  These calculations adopt a bent geometry, similar to the conformation within a biological selectivity filter. This peptide is acetylated at the N--terminus and methylated at the C--terminus to ensure electrostatic neutrality and to emulate a continuous protein backbone, see Fig.~\ref{fig:DFTpointcharges}c.

%\newpage

\subsection{ Electrostatic potential fitting: Extended system}
\par
\begin{figure}[t]
\centering
\includegraphics[width=0.8\textwidth]{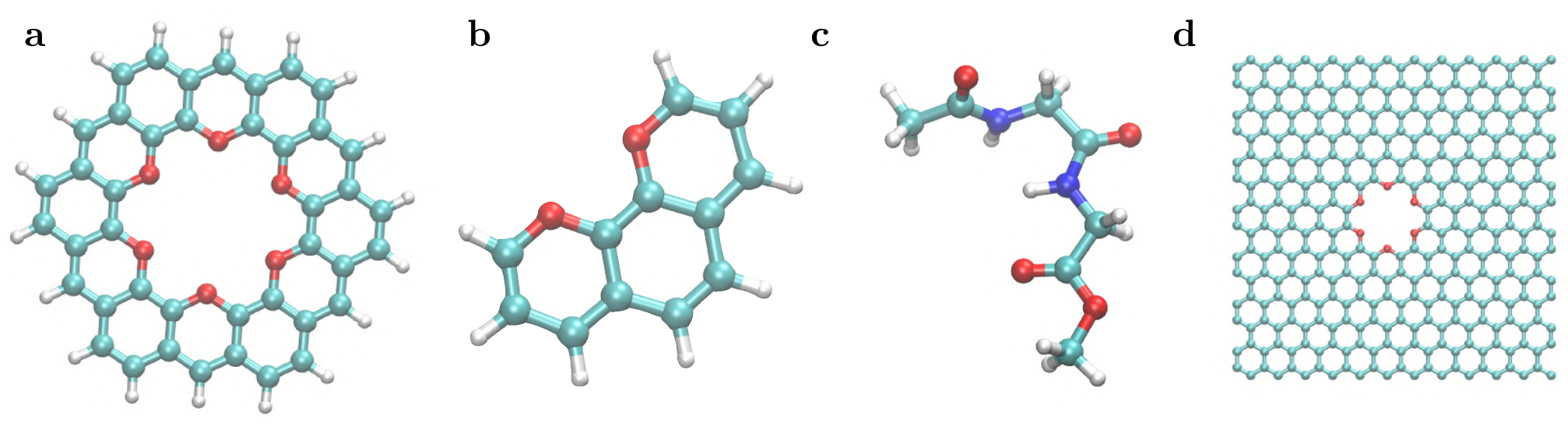} 
\caption{Models employed for (a) CHELPG charge determinations with localized basis methods (B3LYP), (b) CHELPG determinations with correlated methods (RI-MP2), (c)  model CHELPG determinations for a bent  AcN-Gly-Gly-CMe peptide (B3LYP) , and (d)  RESP charge determinations with periodic DFT (PBE-D3).\label{fig:DFTpointcharges}}
\end{figure}

 We also use a complimentary set of plane wave calculations on an 18-crown-6 graphene pore, Fig.~\ref{fig:DFTpointcharges}d, employing a mixed-basis Gaussian plane--wave method in the CP2K code \cite{Hutter2014}. Electronic structure is determined through direct diagonalization of a dispersion-corrected PBE density functional (PBE-D3), alongside GTH pseudopotentials, a localized DZVP basis set, Fermi-Dirac smearing of occupancies ($T=300.0$ K), and a maximal real--space grid cutoff of 5442 eV over four grids~\cite{perdew_generalized_1996,Grimme2010,VandeVondele2005,Godbout1992,Krack2005}. Cell and geometry relaxations use $\Gamma$-point sampling, yielding an optimized supercell measuring 2.957 nm $\times$ 2.988 nm with a 2.0{00} nm vacuum layer perpendicular to the graphene plane. Point charges are calculated using the Restrained Electrostatic Potential Fitting (RESP) algorithm, with sampling optimized for a periodic slab geometry, Table~\ref{tab:charges}~\cite{Bayly1993}.

In order to apply strain, strong binding to a substrate (e.g., a metal) will be required. In a typical nanopore setup, there will be a window approximately 100 nm wide with the graphene overtop. The strong binding interactions, therefore, will be far away from the pore. Since breaking local charge neutrality is energetically unfavorable, binding to the substrate only slightly shifts local electron density. To ensure that this change will not influence the local electrostatic potential felt by a translocating ion (i.e., within the computational approach, this means influencing the assigned partial charges), we examine the effect of additional electron density (e.g., 1~$e$ per 8.84 nm$^2$) and the presence of a perpendicular electric field (up to 5 V/nm). These are well beyond values expected for a practical device, yet neither influence the carbon-oxygen dipole substantially. The excess charge distributes itself more or less uniformly, but with a slight preference for the oxygens. Thus, the setup required for the application of strain should not affect the development of dipoles near the pore, nor should it influence the ion transport other than through the strain

\begin{table}[h]
\centering 
\def\arraystretch{1.2}
\setlength\tabcolsep{8 pt}
\begin{tabular}{ |c|c|c|}
\hline 
Charge  Fitting & Computational Method  & Oxygen Charge  ($q_{\text{O}}$)   \tabularnewline
%Fitting & Method  & ($q_{\text{O}}$)  \tabularnewline
\hline 
\hline 
CHELPG  & DFT/B3LYP  & -0.24$\,e$   \tabularnewline
CHELPG  & RI-MP2  & -0.26$\,e$   \tabularnewline
RESP  & DFT/PBE  & -0.24$\,e$   \tabularnewline
BADER  & DFT/PBE  & {-1.05}$\,e$ %
{}  \tabularnewline
\hline 
\end{tabular}\caption{Oxygen point charges for a 18-crown-6 graphene pore using electrostatic potential fitting (CHELPG, RESP) and Bader analysis. }
\label{tab:charges} 
\end{table}

\section{Molecular Dynamics Simulations}

\subsection{Simulation parameters}
The unique nature of our pore simulations -- specifically the presence of an applied strain -- necessitates careful examination of the standard force-field parameters from CHARMM. A particularly notable parameter is the equilibrium carbon--carbon bond length in graphene, for which a value of $r_{\mathrm{CC}}$ = 137 pm is often taken when performing CHARMM--based molecular dynamics (MD) simulations. Our DFT calculations indicate that this length is inconsistent with an 18-crown-6 pore, see Table~\ref{tab:VASP}, and instead give a value of $r_{\mathrm{CC}}$ = 142 pm. To maintain consistency with electronic structure--based strain models, we use a specific parameter set for graphene~\cite{Feller2000,Smolyanitsky2014molecular}, see Table~\ref{tab:params}. Other parameters are directly from the CHARMM27 force field, and water is taken as TIP3P~\cite{jorgensen1983,Beglov1994}. To enhance stability, we ignore the dihedral energy terms and carbon--carbon Lennard--Jones interactions for atoms within the highly-rigid graphene sheet.

An additional parameter is given by the partial charges $q_{\text{O}}$ on the pore oxygen atoms. Counter--charges must be specified on nearby graphene carbon atoms to maintain charge neutrality. As such, we take the 12 pore--flanking carbons to have a partial charge of $q_{\text{C}}=-q_{\text{O}}/2$, consistent with DFT results. All other carbon atoms are neutral.

\begin{table}[h]
\centering %
\def\arraystretch{1.2}
\setlength\tabcolsep{8 pt}
\begin{tabular}{ |c| c| c| c| c| c|}\hline 
Bond  & $x_{0}$ (nm)  & $k_{b}$ (N/m) &  Angle  & $\theta_{0}$ (nm)  & $k_{a}$ (eV) \tabularnewline\hline 
C--C    & 0.1420  & 698.129  & C--C--C  & 120$^{\circ}$  & 8.008    \tabularnewline
C--O   & 0.1398  & 627.594 & C--C--O  & 120$^{\circ}$  & 3.487     \tabularnewline
      &       &        & C--O--C  & 121$^{\circ}$  & 3.487  \tabularnewline\hline 
\end{tabular}\caption{Force-field parameters to calculate the total bonded energy $E_{\text{tot}}=E_{\text{bond}}+E_{\text{angle}}$ of an 18-crown-6 graphene pore, where $E_{\mathrm{bond}}=\frac{k_{b}}{2}(x-x_{o})^{2}$ is the energy between covalently bonded pairs and $E_{\mathrm{angle}}=\frac{k_{a}}{2}(\theta-\theta_{o})^{2}$ is the angular counterpart.}
\label{tab:params} 
\end{table}

\subsection{Atomistic model setup}

Our MD simulations employ a DFT--optimized graphene sheet with rigid TIP3P water and 1 mol L$^{-1}$ of KCl on both sides. For a membrane of cross--sectional length $L$ nm and height $h_{p}$, the initial height after padding with water is ($1.2L+h_{p}+1$) nm. The factor of 1.2 arises from the `golden aspect ratio' for ion transport simulations, which captures both access and pore resistance~\cite{sahu2018maxwell,sahu2018golden}. \textit{ While we expect access resistance to be less than a 10~\% correction for some parameter regimes, it must be taken into account when understanding the approach to the drift-diffusion limit. Its exclusion by choosing inappropriate (either too small or with a poor aspect ratio) simulation cells hinders the comparison with the drift-diffusion limit.}
The extra 1 nm accommodates packing during equilibration. The construction of simulation cells uses the Visual Molecular Dynamics (VMD) tool~\cite{humphrey1996vmd}. The cross-sectional lengths of the graphene sheet and the pore radius at various strain are given in Table~\ref{tab:radius}.

\begin{table}[h]
\centering %
\def\arraystretch{1.2}
\setlength\tabcolsep{8 pt}
\begin{tabular}{ |c| c| c| c | c| c|}
\hline 
Cell Strain (\%)  & Pore Change (\%)  & $\ell_{x}$ (nm)  & $\ell_{y}$ (nm)  & $r_{n}$ (nm)& $r_{p}$ (nm) \tabularnewline
\hline 
0.0  & 0.0  & 4.650  & 4.521  & 0.290 & 0.137 \tabularnewline
0.5  & 1.0  & 4.673  & 4.538  & 0.292 & 0.140  \tabularnewline
1.0  & 2.1  & 4.696  & 4.566  & 0.295 & 0.143   \tabularnewline
2.0  & 4.8  & 4.743  & 4.611  & 0.303 & 0.151   \tabularnewline
\hline 
\end{tabular}\caption{Pore strain, supercell edge lengths ($\ell_{x},\ell_{y}$), nominal pore
radii ($r_{n}$) and geometric pore radii ($r_p$) as a function of supercell strain. The nominal pore radius $r_n$ is the distance between the center of pore and the center of the edge oxygen atoms. The geometric pore radius is given as $r_p=r_n-r_O$, where $r_O=0.152$ nm is the van der Waals radius of an oxygen atom. The effective pore radius that gives the actual area available for ion transport can be even smaller than $r_p$. This effective radius is \textit{ contextual} and accounts for the totality of interactions (ion--pore, hydration, etc.). It requires analyzing the scatter of translocation events to determine the distance (from the pore center) at which translocations do not occur, see Ref.~\cite{sahu2018maxwell}.}
\label{tab:radius} 
\end{table}

\subsection{Simulation methods}

We perform  all molecular dynamics simulations using the NAMD2~\cite{phillips2005} simulation package, employing a velocity Verlet algorithm to integrate the equations of motion in a fully periodic cell (timestep $\delta t=1$ fs). To expedite simulation, we use a cutoff of 1.2 nm for non--bonded interactions (Lennard--Jones and Coulomb), although a full electrostatic calculation is done every four timesteps using the Particle Mesh Ewald (PME) scheme~\cite{darden1993}.

We initialize simulations with 4000 steps of energy minimization, followed by 4 ps of equilibration in the canonical (NVT) ensemble using a Langevin thermostat (target $T=295$ K, damping $\gamma$ = 0.2 ps$^{-1}$ on heavy atoms). The second round of equilibration is done in the isobaric--isothermal ensemble (NPT), enforced by the Noose--Hover--Langevin piston
(target $P=1.01325\times10^{5}$ Pa) for a total duration of 0.5 ns~\cite{Martyna1994}. We ultimately return to the NVT ensemble for a further 0.5 ns equilibration and production runs. The production runs for the ionic current consist of long NVT simulations (250 ns to 500 ns), depending on convergence, with an electric field along the $z$ direction. Langevin damping is on only the oxygen and carbon atoms, but not on the ions and hydrogen atoms.

Free energy profiles $\Delta F_K (z)$ for the interaction of a \K with the 18-crown-6 pore are determined using the adaptive biasing force (ABF) method~\cite{Henin2004}. We employ a collective variable $z$ -- defined as position along an axis normal to the pore center --\textendash{} that is sampled in a cylindrical region of radius 0.1 nm and length 3 nm for the free energy determination.

\section{Equilibrium free energy and many-body effects}

The net change in energy of a translocating \K is approximately the sum of dehydration energy ($\Delta E_{\mathrm{deh}}$) and several electrostatic terms. The latter component includes terms that give the interaction of the \K with the pore oxygen atoms ($E_\mathrm{KO}$), pore carbon atoms ($E_\mathrm{KC}$), and, if present, a \K that is already in the pore ($E_\mathrm{KK}$). This expression has the form

\begin{align}
\Delta F_{\mathrm{K}}(z) & =\Delta E_\mathrm{deh}+E_\mathrm{KO}+E_\mathrm{KC}+E_\mathrm{KK}\nonumber \\
 & \approx \eta\sum_{i}f_{i}E_{i}+\frac{q_{\mathrm{K}}}{4\pi\epsilon_{0}}\left[\frac{6\,q_{\mathrm{O}}}{\epsilon(z)\sqrt{z^{2}+r_{O}^{2}}}+\frac{12\,q_{\mathrm{C}}}{\epsilon(z)\sqrt{z^{2}+r_{C}^{2}}}+\frac{q_{\mathrm{K}}}{\epsilon(d_{\mathrm{KK}})\,d_{\mathrm{KK}}}\right],
\end{align}
where $f_{i}$ and $E_{i}$ are the fractional dehydration and the energy corresponding to the $i^{\mathrm{th}}$ hydration layer, respectively; the factor $\eta\approx 1/2$~\cite{sahu2017dehydration} is due to stronger orientation of water dipoles remaining in the hydration layer after some have been lost;  $\epsilon(z)$ is a position-dependent relative permittivity; $q_{\mathrm{\nu}}$ is the charge of ionic species $\nu$; and $d_{\mathrm{KK}}$ is the distance between a \K and a \K that may be present in the pore.

\begin{figure}[h]
\centering
\includegraphics[width=\textwidth]{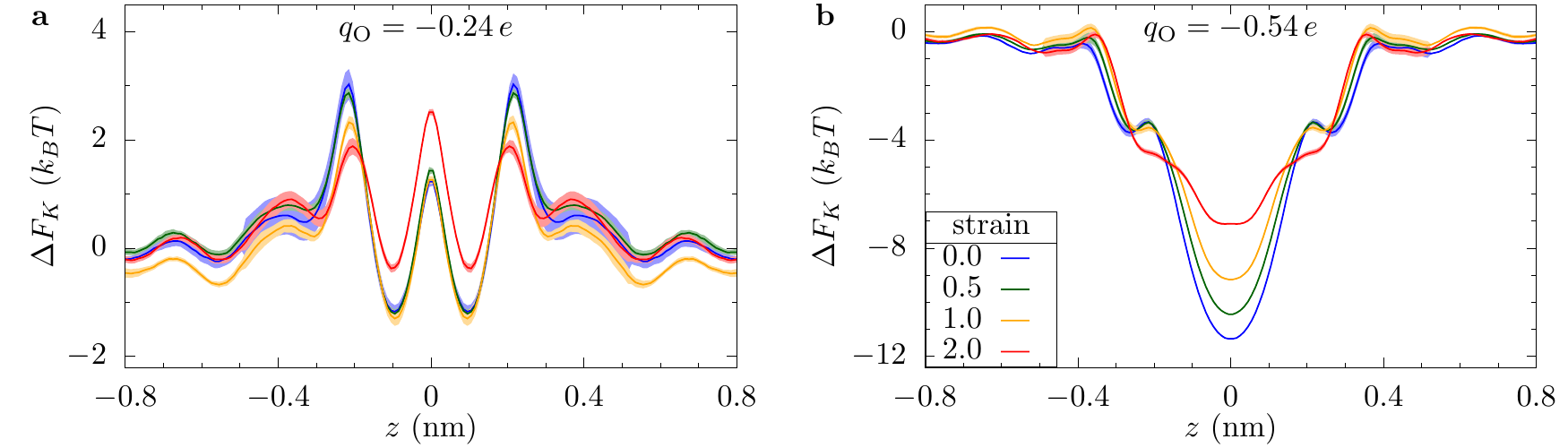}
\caption{Free energy profile when only one \K and one \Cl are present in the solution. (a) The free energy profile for $q_\mathrm{O}=-0.24\,e$ remains almost identical to the case for a 1 mol/L solution, as described within in the main text. (b) For $q_\mathrm{O}=-0.54\,e$, however, the free energy profile exhibits significant differences. In the absence of the possibility of KK interaction, there is no barrier outside the pore, and the potential well at the pore center becomes significantly deeper due to the lack of charge screening due to other \K in solution. These free-energy profiles show that the intermediate level of charge is in a many-body regime, whereas the smaller pore charge is mostly single ion physics.\label{fig:pmf1KCl}}
\end{figure}

For $q_\mathrm{O}=-0.24\,e$, the $E_\mathrm{KK}$ term doesn't play a significant role because the translocating \K generally finds the pore empty. The free energy profile when only a single \K and a counterion (Cl$^-$) are present in the whole simulation cell is shown in Fig.~\ref{fig:pmf1KCl}.  For $q_\mathrm{O}=-0.24\,e$, this is similar to the 1 mol/L solution described in the main text. This profile changes markedly when $q_\mathrm{O}=-0.54\,e$. In this intermediate--charge case, there is no free energy barrier outside the pore, indicating that the barrier seen for a 1 mol/L solution is due to KK interaction.  The well depth also decreases substantially, as the solution is now less effective at screening the pore rim and there is no repulsion from other cations, both of which allow the interstitial \K to bind more strongly.

\begin{figure}[h]
\centering
\includegraphics[width=0.45\textwidth]{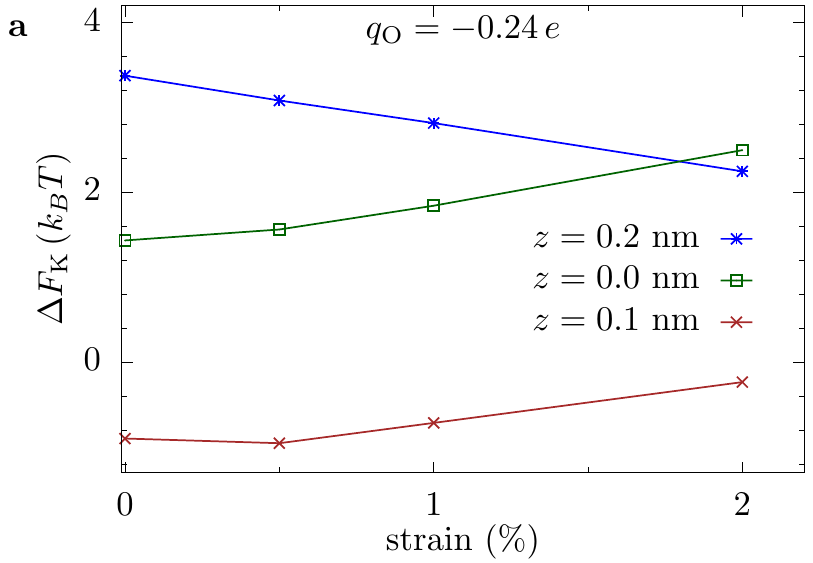}\quad
\includegraphics[width=0.48\textwidth]{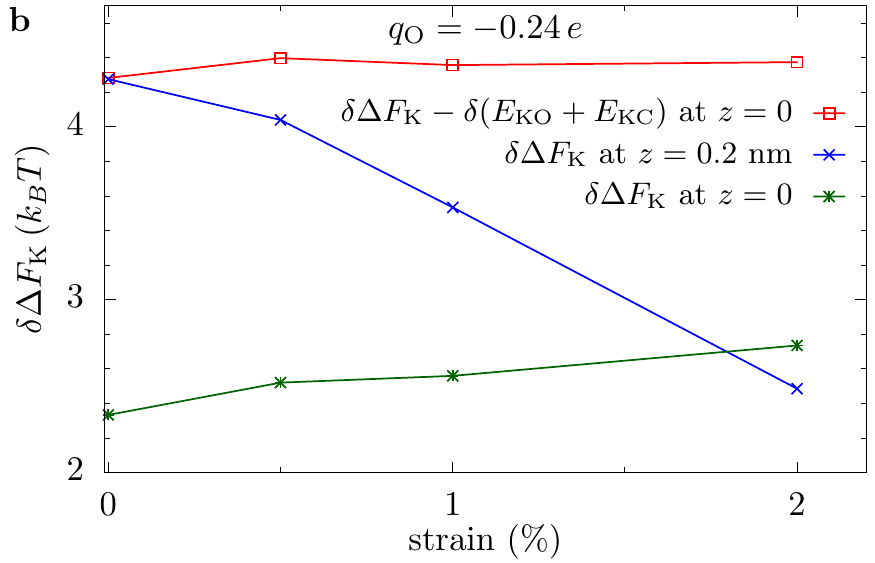}
\caption{Free energy dependence on strain for $q_\mathrm{O}=-0.24\,e$ near pore. (a) Free energy peaks and wells at different positions versus strain. The outer peaks, $\Delta F_\mathrm{K}(z=\pm 0.2\,\text{nm})$, decrease with strain, whereas the central peak $\Delta F_\mathrm{K}(z= 0)$ and neighboring wells $\Delta F_\mathrm{K}(z=\pm 0.1\,\text{nm})$ increase with strain. These increases go nearly in tandem, with the wells increasing a bit more slowly because they are further from the charged-groups on the pore rim (and thus the electrostatic effect is slightly weaker). (b)  Free energy barrier measured from the well bottom at $z=0.1$ nm to the peaks at $z_2=0.2$ and $z_2=0$ nm, $\delta \Delta F_\mathrm{K}=\Delta F_\mathrm{K}(z_2)-\Delta F_\mathrm{K}(z_1)$. For small strain, the maximum barrier is the outer barrier (blue) connecting the external solution and either of the the potential minima ($z = \pm 0.1$ nm). This decrease with strain is mostly due to an increase in hydration energy. The barrier between $z=0$ and $z=0.1$ nm (green) increases with strain due the change in electrostatic energy. The contribution from dehydration alone (red), $\delta \Delta F_\mathrm{K}-\delta(E_\mathrm{KO}+E_\mathrm{KC})$  at $z=0$, is fairly insensitive to strain.\label{fig:deltaF0.24} }
\end{figure}

Three free energy barriers ($\delta \Delta F_\mathrm{K}=\Delta F_\mathrm{K}(z_2)-\Delta F_\mathrm{K}(z_1)$) are present when  $q_\mathrm{O}=-0.24\,e$.  The central barrier is situated between $z_2=0$ nm and $z_1=\pm 0.1$ nm, accompanied by two partners between $z_2=\pm 0.2$ nm and $z_1=\pm 0.1$~nm. The barrier at $z=0$ is relatively insensitive to strain, as dehydration in the pore region is not affected by the small deformations that we consider. The small increase in the $\delta \Delta F_\mathrm{K}$ is due to a change in electrostatic energies ($E_\mathrm{KO}$ and $E_\mathrm{KC}$). The barrier due to dehydration alone $\delta \Delta F_\mathrm{K}-\delta(E_\mathrm{KO}+E_\mathrm{KC})$ remains fairly constant as shown in Fig.~\ref{fig:deltaF0.24}.

We note that the change in the free energy barrier between systems with 1 mol/L KCl (or any concentration) and those containing only 1 \K and 1 \Cl indicates the presence of many-body effects, as we have discussed above. In the case of $q_\text{O} = -0.54\,e$, the external barrier (at $z \approx 0.2$~nm) disappears when going to just one ion and one counterion in solution. This barrier is due to the presence of another \K in the pore.

\begin{figure}[h]
\centering
\includegraphics[width=0.45\textwidth]{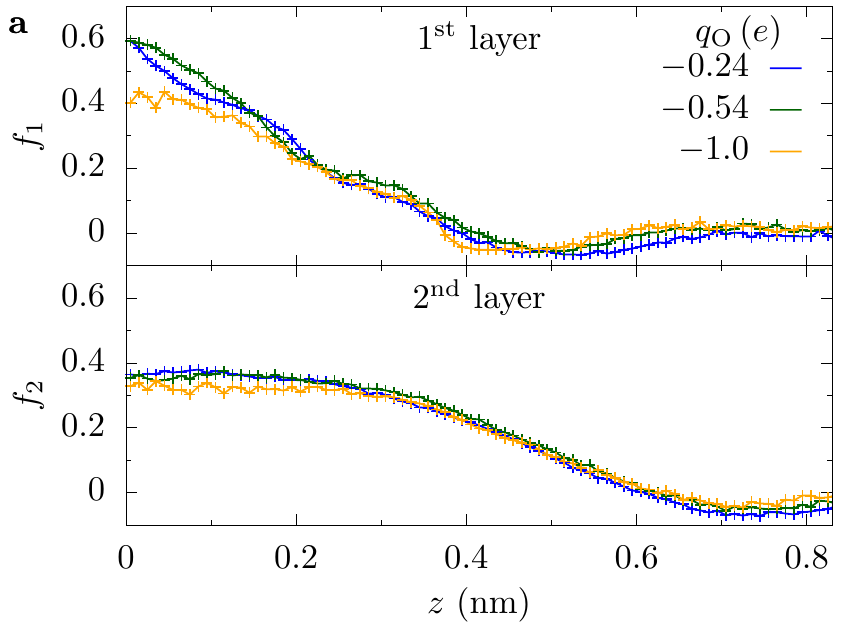}\quad{}
\includegraphics[width=0.45\textwidth]{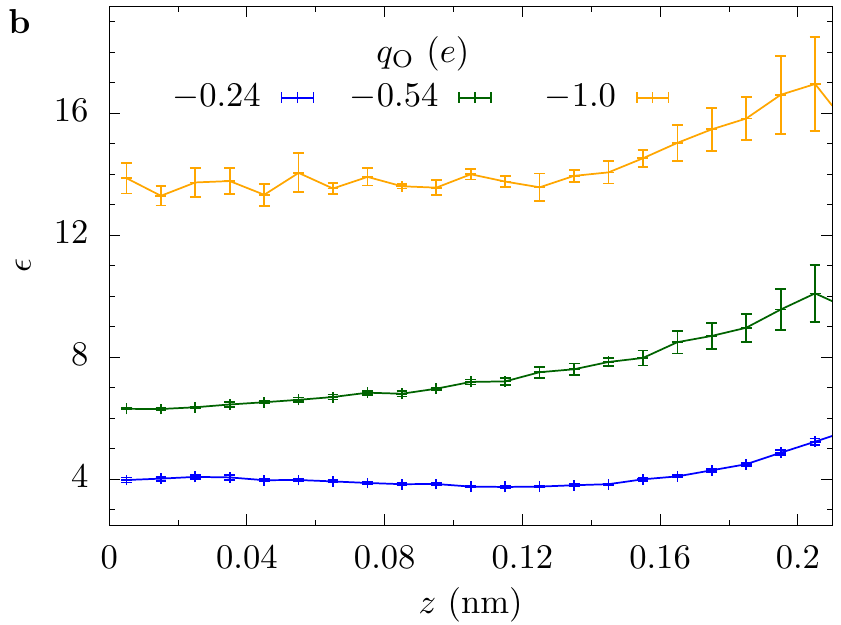}
\caption{Dehydration effects and the dielectric constant in confined geometries. (a) Fractional dehydration in the first, $f_1$, and the second, $f_2$, hydration layer of a \K as it translocates through an 18-crown-6 graphene pore. (b) Relative permittivity $\epsilon(z)$ for different values of $q_{\mathrm{O}}$, as determined using Eq.~\ref{Eq:eps} and MD values for $\Delta F_{\mathrm{K}}$. The magnitude of $\epsilon(z)$ departs markedly from the bulk value ($\epsilon=79$) near the pore. Beyond $|z|>0.2$ nm, $E _{\mathrm{KK}}$ can be significant and thus Eq.~\ref{Eq:eps} cannot not be used to calculate $\epsilon$. Error bars are standard errors from binned data. \label{fig:eps} }
\end{figure}

\vspace{1\baselineskip}
\noindent{\bf Relative Permittivity}\\
When a \K is very close to the pore, we can ignore the $E_\mathrm{KK}$ contribution to the free energy so that   
\begin{align}
\Delta F_{\mathrm{K}} & \approx \eta\sum_{i}f_{i}E_{i}+\frac{q_{\mathrm{K}}}{4\pi\epsilon_{0}\,\epsilon(z)}\left[\frac{6\,q_{\mathrm{O}}}{\sqrt{z^{2}+r_{O}^{2}}}+\frac{12\,q_{\mathrm{C}}}{\sqrt{z^{2}+r_{C}^{2}}}\right].\label{Eq:energy1K}
\end{align}
from which we may adopt a convenient representation for the position--dependent permittivity,
\begin{align}
\epsilon(z) & \approx \frac{q_{\mathrm{K}}}{4\pi\epsilon_{0}(\Delta F_{\mathrm{K}}-\eta\sum_{i}f_{i}E_{i})}\left[\frac{6\,q_{\mathrm{O}}}{\sqrt{z^{2}+r_{O}^{2}}}+\frac{12\,q_{\mathrm{C}}}{\sqrt{z^{2}+r_{C}^{2}}}\right].\label{Eq:eps}
\end{align}

We employ the fractional dehydration $f_{i}=1-n_{\text{water}}^{(i)}(z)/n_{\text{bulk}}^{(i)}$ of \K to estimate $\epsilon(z)$. In Fig.~\ref{fig:eps}a, we plot the fractional dehydration in the first and the second
hydration shell of an ion translocating through the pore, alongside the $\epsilon(z)$  calculated from Eq.~\ref{Eq:eps} in  Fig.~\ref{fig:eps}b. We find that the magnitude of $\epsilon(z)$ is dramatically reduced from its bulk value $\epsilon=79$ as the ion approaches the pore center. A low dielectric constant is expected for the sub--nanometer distances between these charged particles. Farther from the pore, the magnitude of $\epsilon(z)$ will rise as the hydration layers become complete. However, Eq.~\ref{Eq:eps} only allows us to reliably calculate $\epsilon(z)$ very close to the pore

\section{Ion transport mechanism: Knock-on versus drift-diffusion}
\begin{figure}[h]
\centering
\includegraphics{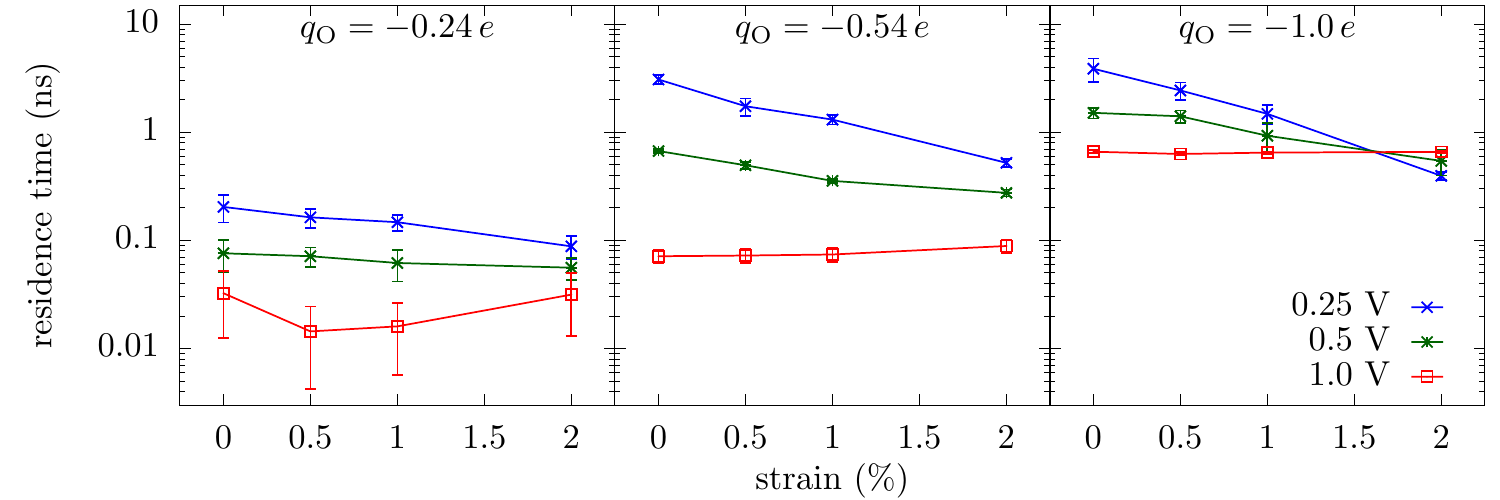}
\caption{Residence time for a \K translocating through crown ether graphene pore in 1 mol/L KCl solution with various values of $q_\mathrm{O}$, strain, and applied voltage. The residence time is short for small $q_\mathrm{O}$, moderate strain, and large voltage, suggesting a drift--diffusion type translocation. For large $q_\mathrm{O}$, small-to-no strain, and small voltage, the residence time is significant, suggesting a knock--on mechanism. At zero bias, the residence time increases by a factor of 2 to 4 for most data points. The residence times for $q_\mathrm{O}=-0.54\,e$ and $q_\mathrm{O}=-1.0\,e$ are strongly influenced by the ion concentration; the residence time with only one K$^+$ and one Cl$^-$ in the simulation cell is over 100 ns. Error bars are $\sqrt{\sigma^2+\Delta T^2}$ where $\sigma$ is the standard error from five parallel runs and $\Delta T=10$ ps is the sampling time. \label{fig:occT} }
\end{figure}

\begin{figure}[h]
\centering
\includegraphics{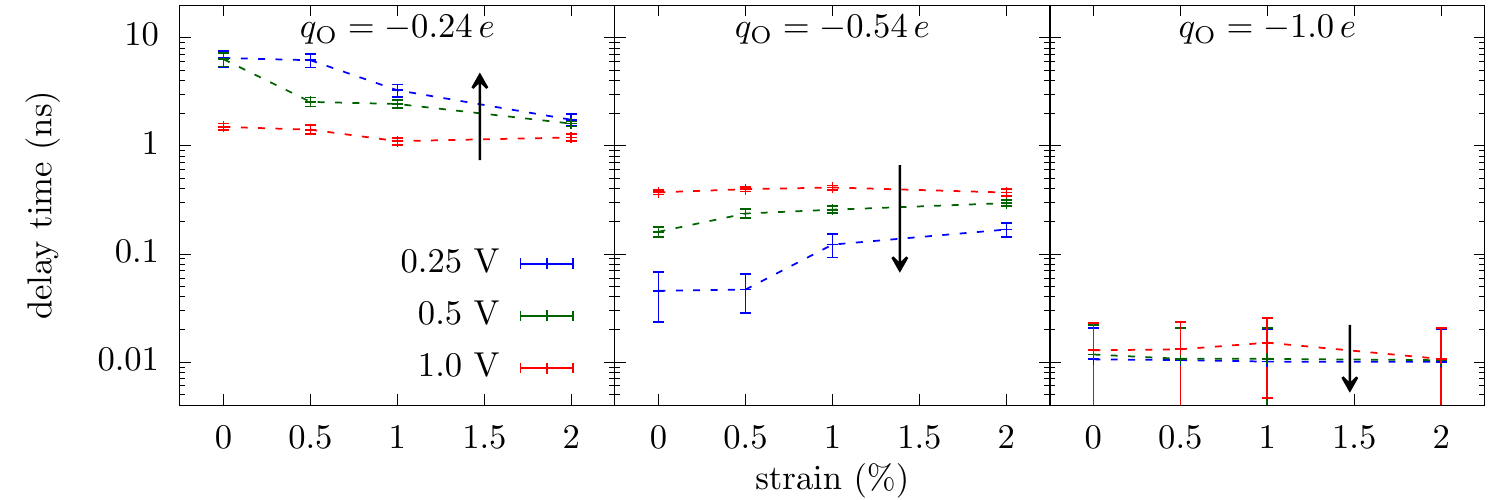}
\caption{Delay time between one \K leaving the pore and another \K replacing it for various values of $q_\mathrm{O}$, strain and applied voltage. In drift--diffusion transport, which occurs for small $q_\mathrm{O}$, moderate strain, and large voltage, the delay time is long and thus the pore is empty most of the time. By contrast, in knock--on type transport, which occurs for large $q_\mathrm{O}$, small-to-no strain, and small voltage, the delay time is short and thus pore is essentially always occupied by a K$^+$. The arrows note that, in the zero bias limit, drift-diffusion and knock-on delay times behave differently: The dichotomy is indicative of the fact that for drift-diffusion the delay is related to the current but knock-on is related to the inability of a trapped \K to leave the pore without assistance. The former thus has a delay that gets longer with decreasing bias, as \K does not stay in the pore and the delay is determined by the magnitude of the current (or, equivalently, the driving force to move ions from bulk into the pore). The latter delay gets shorter with smaller bias, as the pore wants to always be occupied and only the replacement with another \K can remove the interstitial \K (i.e., the bias no longer helps in the process). For both the larger partial charge cases, the delay time is fast (a few ps) for 0 V. Error bars are $\sqrt{\sigma^2+\Delta T^2}$ where $\sigma$ is the standard error from five parallel runs and $\Delta T= 10$~ps is the sampling time. \label{fig:repT} }
\end{figure}

The mechanism of ion transport is either of knock--on or drift--diffusion character, as determined by a combination of charge, strain, and applied voltage.  Qualitatively, the distinction between these mechanisms can be made by directly observing the simulation trajectory. Alternately, the ion residence time (the average time an ion spends in pore) and the delay time (the average time interval between one ion leaving the pore and next ion replacing it) can quantify aspects of the transport mechanism. In knock--on transport, a translocating \K  spends significant time in the pore -- on the order of a nanosecond -- and is replaced by another \K as soon as it departs (see Figs.~\ref{fig:occT} and \ref{fig:repT}). Conversely, in a drift--diffusion scenario, a \K spends less than 0.1 ns in the pore.  The pore will then remain vacant for several nanoseconds after the ion departs. From this data, it is clear that the smallest charge is drift-diffusion and the largest charge is knock-on. The intermediate charge is a mix of the two mechanisms with weights that change as the pore undergoes strain.

\newpage
\section{Ion transport through multiple barriers}

To address transport in the small charge ($q_\text{O} = -0.24\,e$) regime, we model the pore as a central dehydration barrier of height $U$, flanked by two satellite barriers of height $E_B$. We assume that the potential drop $V$ is spatially localized to the graphene interface, uniformly displacing the background by $V/2$  and $-V/2$ on the high-- and low--bias sides of the membrane. This convention leaves satellite barriers invariant while reducing the central barrier to $U - V/2$ for an ion approaching from high bias side and increasing it to $U + V/2$ for permeation at low bias.  

\par  We assume that ions permeate the outer dehydration barriers at a rate $k_\text{in} c$, where $c$ is the bulk \K concentration.   Due to strong repulsion between \K pairs, the pore will be preferentially occupied by a single \K.  The transfer of \K out of the pore is then an activated process, with rate $k = k_\text{out} e^{-E_B/k_B T}$\cite{nelson2002permeation}.  An ion crosses the central barrier through a similar mechanism, with the rate prefactor $k_\text{pore}$ taken as identical on high-- and low--bias sides. Given these considerations, the current through the pore is

\begin{equation}
I = \frac{2 H_1\gamma}{H_2+2\gamma} \frac{[H_2 + 2\gamma]^2(3 + 2\alpha)\alpha - (1+2\alpha)(H_2 + 2\gamma)\sqrt{\alpha((H_2 + 2\gamma)^2(2+\alpha) - 2 H_1^2)}} {H_1^2[1 + 4(1+\alpha)\alpha)] - H_2 (H_2 + 4\gamma) - 4\gamma^2}.
\end{equation}
For notational simplicity, we suppress an explicit dependence on the bias potential and barrier energies.  The central potential and applied bias manifest through a pair of parameters $H_1 (U,V) = 2k_\text{pore}\, e^{-U/k_BT}\sinh (V/2k_B T)$ and  $H_2 (U,V) = 2k_\text{pore}\, e^{-U / k_B T }\cosh ( V/2 k_B T)$, while the satellite barriers contribute through the factors $\alpha (E_B) = (k_\text{out} / 2 k_\text{in}c) e^{-E_B /k_B T}$ and $\gamma (E_B) = k_\text{in}c\,\alpha( E_B)$.  If we approximate all rate prefactors as equal and modulate potentials so that $U = U_0 + \delta U$ and $E_B = U_0 - \delta U$ as $V / U_0 \longrightarrow 0$, we find that $I(V)$ is maximized for $\delta U / k_B T = 0.339 $ when $U_0 / k_B T = 1$, or $\delta U / k_B T = 0.503 $ when $U_0 / k_B T = 2$.  The latter parameters approximate the free energy profile for $q_\text{O} = -0.24\,e$ at 2\% strain. While the parameter space gives a more complicated structure, the optimum regime for this type of modulation captures the behavior in Fig.~4 of the main text.

%\end{document}
%\clearpage

%\bibliography{ref}